\documentclass{article}

\usepackage[T1]{fontenc}
\usepackage[utf8]{inputenc}
\usepackage{arxiv}                 

\usepackage{amsmath,amssymb}
\usepackage{graphicx}
\usepackage{booktabs}
\usepackage[round,authoryear]{natbib}   
\usepackage{hyperref}
\hypersetup{colorlinks=true, linkcolor=blue, citecolor=blue, urlcolor=blue}

\title{How effective normal stress oscillations advance failure in fault gouge: frequency dependence, non-failure window, and the role of dilation}

\renewcommand{\shorttitle}{Effective normal stress oscillations advance failure in fault gouge}

\renewcommand{\undertitle}{}

\author{%
  Pritom Sarma\textsuperscript{1}\thanks{Corresponding author: \texttt{pritom.sarma@mail.huji.ac.il}}\quad
  Einat Aharonov\textsuperscript{1,2}\quad
  Renaud Toussaint\textsuperscript{2,3}\quad
  Stanislav Parez\textsuperscript{4,5} \\[0.6ex]
  \small\textsuperscript{1}Institute of Earth Sciences, Hebrew University of Jerusalem, Jerusalem, Israel \\
  \small\textsuperscript{2}PoreLab, Departments of Physics and Geosciences, The Njord Centre, University of Oslo, Oslo, Norway \\
  \small\textsuperscript{3}Universit\'{e} de Strasbourg, CNRS, ENGEES, Institut Terre et Environnement de Strasbourg, UMR7063, Strasbourg, France \\
  \small\textsuperscript{4}Faculty of Science, Jan Evangelista Purkyn\v{e} University, \'{U}st\'{\i} nad Labem, Czech Republic \\
  \small\textsuperscript{5}Faculty of Science, Charles University, Prague, Czech Republic%
}
\date{}

\begin{document}
\maketitle

\begin{abstract}
Cyclic pore-pressure or normal stress variations arise both in relation to natural earthquakes and in engineered subsurface systems, yet their effect on fault stability remains poorly constrained at the grain scale. Here we numerically model, using a coupled Discrete Element--fluid dynamics model, the response of a sheared, fluid-saturated or dry, gouge-filled fault to effective normal stress oscillations over a wide frequency range (0.5--$10^4$~Hz). The effective normal stress is oscillated either by cycling the pore-pressure or by directly cycling the normal stress, while keeping the stress state below the Mohr--Coulomb threshold measured in continuous loading. Despite this sub-critical loading, we observe failure across most frequencies, with a non-monotonic frequency dependence. A distinct non-failure window emerges at intermediate frequencies ($\sim$30--200~Hz), bounded by failure at both lower and higher frequencies; the system exhibits four regimes from cyclic failure-and-arrest to continuous sliding. Pore-pressure and normal stress oscillations produce the same regime structure, confirming that they act as equivalent forcings via Terzaghi's principle, with fluid coupling adding only a delay due to dilatant hardening. Sub-critical failure arises from dilation-induced strength deterioration via two mechanisms: (i) low-frequency cycles allow sufficient time for shear-driven ratcheting dilation, while (ii) high-frequency cycles induce dynamic dilation (acoustic fluidization) via amplified seepage forces, stress gradients and inertial forces. The intermediate non-failure window represents the gap between these mechanisms. These results identify frequency as a controlling parameter for failure in granular materials, with implications for dynamic earthquake triggering and cyclic injection protocols.
\end{abstract}


\noindent\textbf{Highlights}
\begin{itemize}\setlength{\itemsep}{0pt}
\item Cyclic effective normal stress drives sub-Coulomb failure of fault gouge
\item Three failure modes and a non-failure window emerge with frequency.
\item Shear-driven ratcheting dilation leads to failure at low frequencies.
\item Acoustic fluidization (AF) leads to failure at high frequencies.
\end{itemize}

\section{Introduction}
\label{sec:intro}

Active geological faults are continually subjected to time-varying loads that modulate their strength and stability. A central control on fault strength is the effective normal stress $\sigma'_n = \sigma_n - P$, the difference between the total fault-normal stress $\sigma_n$ and the pore-fluid pressure $P$ \citep{Terzaghi1996, Hubbert1959}. Both quantities fluctuate in time, in nature and in engineered settings: $\sigma_n$ is perturbed by solid-Earth and ocean tides, by the transient stresses carried by passing seismic waves, and by surface and machinery loading, whereas $P$ is varied by fluid injection and withdrawal, by tidal pore-pressure changes, by tectonic and hydrological processes. Because failure is governed by $\sigma'_n$ and not by either term in isolation, lowering $\sigma_n$ and raising $P$ are mechanically equivalent routes to weakening. How a fault responds to \emph{cyclic} variations of $\sigma'_n$, imposed through either route, is therefore a problem of fundamental importance for earthquake physics and for the safe management of subsurface operations. Faults in the Earth's crust frequently host a layer of granular material known as fault gouge, formed by wear and comminution of the host rock during repeated episodes of slip \citep{Scholz2019}. These gouge layers are typically fluid-saturated, and their frictional strength, stability, and propensity for failure control the seismic behavior of the host fault \citep{Kanamori2004, Marone1998, Marone1990, Dorostkar2017, Scuderi2015, Scuderi2022, Zhang1998, Collettini2022}. Increasing pore pressure lowers the shear resistance $\tau = \mu(\sigma_n - P)$, where $\mu$ is a Coulomb friction coefficient; this underpins the link between subsurface fluid injection (enhanced oil recovery, geothermal extraction, wastewater disposal, and CO$_2$ sequestration) and induced seismicity, commonly attributed to pore-pressure diffusion to pre-stressed faults \citep{Ellsworth2013, Keranen2018, Schultz2020}.

While most studies of injection-induced seismicity consider monotonic or constant-rate pore-pressure loading \citep{Cappa2019, Guglielmi2015, Sarma2025}, many natural and industrial scenarios involve \textit{cyclic} variations of the effective normal stress. Earth and ocean tides produce periodic pore-pressure and effective-stress variations \citep{Cochran2004, Thomas2009} linked to modulation of seismicity rates \citep{Metivier2009, vanderElst2010}; seismic waves trigger earthquakes on distant faults across a range of frequencies \citep{Gomberg2005, HillPrejean2007}; and tidal loading of ocean floors and ice sheets modulates seismicity in subduction and glacial settings \citep{Rubinstein2008}. In industry, pulsed or oscillatory injection has been explored to manage induced seismicity in geothermal reservoirs \citep{Zang2019, Kwiatek2019, Hofmann2018}.

A substantial body of experimental, theoretical, and numerical work has established how faults respond to oscillations of the normal stress. The foundations are in the dry, normal-stress-driven case. Variations in $\sigma_n$ couple directly into the state evolution of rate-and-state friction and can destabilize sliding that would otherwise be stable, even where the classical critical-stiffness criterion predicts stability \citep{Dieterich1992, Linker1992, Hong2005}. Building on this, friction experiments on simulated gouge have shown that periodic $\sigma_n$ oscillations weaken the layer below its steady-state Coulomb strength, with the weakening growing as the oscillation amplitude increases and the period shortens, and with the response sensitive to grain size and effective stress \citep{Boettcher2004, Yu2024}. Large, rapid oscillations can trigger slow laboratory earthquakes even while the shear stress stays below the Coulomb threshold, reproducing the range of slip behaviours observed on natural faults \citep{Pignalberi2024}. Rotary-shear and in-situ experiments link such rapid $\sigma_n$ oscillations to permanent (anelastic) dilation and weakening of the gouge, although whether their net effect is to stabilize or destabilize the fault is still debated \citep{Chen2024}. Closely related transient-stress and shear-velocity perturbation experiments confirm sub-Coulomb yielding on quartz and granite gouges \citep{Savage2007}, and dynamic-triggering experiments on sheared glass-bead layers show that small acoustic perturbations trigger slip below the steady-state Coulomb criterion \citep{Johnson2008}. 

Grain-scale Discrete Element Method (DEM) studies provide a complementary view: boundary vibration above a threshold amplitude advances the next slip event in a sheared granular layer through non-affine grain rearrangement, with a sharp, first-order-like onset above a critical strain amplitude of order $10^{-6}$ \citep{Griffa2013, Ferdowsi2015}, and the resulting frequency response is consistent with sub-critical failure driven by perturbations to the friction state variable. Together, this literature establishes that cyclically-loaded faults can fail well below the static Mohr--Coulomb envelope. What has not been systematically addressed is (a) the response across a wide frequency range, including conditions under which the system does \emph{not} fail; (b) the underlying grain-scale mechanism for the frequency dependence; and (c) whether pore-pressure oscillations on a saturated layer produce the same response as direct normal-stress oscillations on a dry layer.

A critical ingredient in granular failure is shear-induced dilation. In dry granular media it has long been established that a densely packed layer must dilate in order to shear \citep{Reynolds1885, Reynolds1886, Mead1925}, and that, because work is done against the normal stress as grains climb over one another, the apparent friction of the layer decreases as its porosity increases \citep{Rowe1962, Marone1990, FryeMarone2002, Bolton1986, Makedonska2011, Chen2016}. The dependence of friction on porosity, $d\mu/d\phi < 0$, is thus a basic and general property of sheared granular materials. A convenient analytical expression for this dependence is the microphysical grain model of \citet{Chen2016},
\begin{equation}
\tau = \mu(\phi)(\sigma_n - P),
\label{eq:tau_mu}
\end{equation}
with
\begin{equation}
\mu(\phi) = \frac{\mu_s^c + 2H(\phi_c - \phi)}{1 - 2\mu_s^c H(\phi_c - \phi)},
\label{eq:chen_spiers}
\end{equation}
where $\mu_s^c$ is the static friction coefficient of grain contacts, $H$ a geometric constant, $\phi_c$ the critical-state porosity, and $\phi$ the current porosity; since $d\mu(\phi)/d\phi < 0$, increasing porosity weakens the layer. Equation~\ref{eq:chen_spiers} is not an exact law: it is derived for uniform spherical grains, and no closed-form $\mu(\phi)$ exists for arbitrary (polydisperse, angular, three-dimensional) packings \citep{Makedonska2011}. We use it only as a concrete illustration of the general relation $d\mu/d\phi < 0$, in the pre-failure quasi-static regime where $\mu(\phi_c)$ is the lowest (residual) friction the contacts can reach; it is not expected to hold once the layer accelerates into the inertial regime, where instantaneous porosity can exceed $\phi_c$ and friction follows inertial-number ($\mu(I)$) rheology \citep{daCruz2002, MiDi2004, Forterre2008}. The critical-state porosity is also not fixed: for sheared, fluid-saturated gouge it depends on effective stress and, more weakly, on sliding velocity and the evolving grain fabric \citep{Segall1995, Pailha2008}.

In fluid-saturated systems, dilation additionally produces a transient pore-pressure reduction (dilatant hardening) that temporarily stabilizes the layer \citep{Segall1995, Frank1965}; the competition between dilation rate and pore-pressure diffusion then controls the onset of failure and introduces a delay between loading and rupture \citep{Pailha2008, Pailha2009, Sarma2025}. \citet{Sarma2025} demonstrated using coupled fluid-DEM simulations that fluid-saturated fault gouge fails through preparatory dilation, dilatant hardening, and competition between porosity evolution and pore-pressure diffusion.

Here we use a coupled fluid-DEM model \citep{Cundall1979, Aharonov1999, Goren2011, BenZeev2023, Parez2023, Sarma2025} to investigate the response of a granular fault gouge layer to cyclic effective normal stress oscillations from 0.5 to $10^4$~Hz, using two complementary protocols: (1) direct cyclic $\sigma_n$ oscillations on a dry layer (control), anchored to the established literature \citep{Linker1992, Boettcher2004, Pignalberi2024}; and (2) cyclic pore-pressure oscillations of equivalent effective-stress amplitude on a fluid-saturated layer, testing whether fluid coupling modifies the response. Peak loading always remains below the Mohr--Coulomb criterion. We identify three failure modes plus a non-failure window and explain the frequency dependence through (i) shear-induced dilation, (ii) the competition between momentum-diffusion timescale and oscillation period, and (iii) frequency-dependent stress gradients and inertial forces, providing a grain-scale explanation for sub-Coulomb failure under oscillatory loading, with direct implications for dynamic earthquake triggering and cyclic injection protocols in geothermal energy and CO$_2$ sequestration.

\section{Methods}
\label{sec:methods}

\subsection{Numerical scheme: coupled solid-fluid DEM formulation}
\label{sec:numerical_scheme}

We simulate grain-level micromechanics within a sheared granular fault gouge using a two-dimensional Discrete Element Method (DEM) \citep{Cundall1979} coupled with a continuum fluid solver for the pore fluid pressure. The coupled model, developed and validated in previous studies \citep{McNamara2000, Goren2010a, Goren2011, BenZeev2023, Parez2023, Sarma2025}, is summarized here.

\subsubsection{Fluid phase}

The pore-fluid mechanics follows the formulation of \citet{Goren2010a, Goren2011}. Mass conservation for grains and fluid is
\begin{equation}
\frac{\partial[(1-\phi)\rho_s]}{\partial t} + \nabla \cdot [(1-\phi)\rho_s \mathbf{u_s}] = 0,
\label{eq:mass_solid}
\end{equation}
\begin{equation}
\frac{\partial[\phi \rho_f]}{\partial t} + \nabla \cdot [\phi \rho_f \mathbf{u_f}] = 0,
\label{eq:mass_fluid}
\end{equation}
where $\rho_s$, $\rho_f$ are solid and fluid densities, $\mathbf{u_s}$, $\mathbf{u_f}$ are the solid and fluid velocity fields, and $\phi$ is porosity. Fluid momentum balance equation is approximated by Darcy's law,
\begin{equation}
\phi(\mathbf{u_f} - \mathbf{u_s}) = -\frac{k}{\eta}\nabla P,
\label{eq:darcy}
\end{equation}
where $P$ is the excess fluid pressure above the hydrostatic level, $k$ is permeability, and $\eta$ is fluid dynamic viscosity. A linearized state equation $\rho_f = \rho_0(1 + \beta P)$ closes the system, with $\beta$ the adiabatic fluid compressibility. The solid phase is treated as incompressible. Combining these yields the governing equation for excess pore pressure \citep{Goren2011, Parez2023}:
\begin{equation}
\frac{\partial P}{\partial t} - \frac{1}{\beta \phi \eta}\nabla\cdot[k\nabla P] + \frac{1}{\beta \phi}\nabla \cdot \mathbf{u_s} = 0,
\label{eq:pore_pressure}
\end{equation}
where $\nabla \cdot \mathbf{u_s}$ is the local volumetric strain rate (negative for compaction, positive for dilation). The derivation of Equation~\ref{eq:pore_pressure} neglects an advective term, $\beta \phi\,\mathbf{u_s}\!\cdot\!\nabla P$, which represents transport of pore pressure by the moving grain skeleton; this term is negligible whenever the pore-pressure diffusion length remains larger than a grain diameter, so that pressure transport is diffusion-dominated rather than advective \citep{Goren2010a, Goren2011}. Equation~\ref{eq:pore_pressure} is solved on a regular grid with spacing equal to two mean grain diameters; $k$, $\phi$, $\mathbf{u_s}$ are interpolated from grains onto the grid using bilinear interpolation \citep{Goren2011}. Permeability and porosity are related through a modified Carman--Kozeny relationship \citep{McNamara2000}:
\begin{equation}
k = k_c \frac{(1 + 2\phi_{2D})^2}{(1-\phi_{2D})^2},
\label{eq:permeability}
\end{equation}
where $k_c$ is a permeability prefactor and $\phi_{2D}$ is the 2D porosity.

\subsubsection{Solid phase}

The solid phase is modeled using DEM with a linear elastic frictional contact model \citep{Cundall1979}. Grains are represented as 2D disks interacting via pairwise contact forces. Linear and rotational momentum conservation are
\begin{equation}
m_i \dot{\mathbf{u}}_{s,i} = \sum_j \mathbf{F}_{ij} - \frac{V_i}{1-\phi}\nabla P,
\label{eq:linear_mom}
\end{equation}
\begin{equation}
I_i \dot{\omega}_{s,i} = \sum_j R_i \hat{n}_{ij} \times \mathbf{F}_{ij},
\label{eq:angular_mom}
\end{equation}
where $m_i$, $I_i$, $V_i$, $R_i$ are grain mass, moment of inertia, volume, and radius. The first term in Eq.~\ref{eq:linear_mom} is the sum of contact forces; the second is the drag force from the pore-pressure gradient. Equations are time-integrated using the Verlet algorithm. The contact force $\mathbf{F}_{ij}$ uses a linear spring with velocity-dependent damping and the Coulomb friction criterion (see \citep{Parez2023}, Eq.~23 for tangential displacement evolution). The two-phase code was validated against poro-elastic compression experiments \citep{Nur1971} and granular-fluid instability experiments \citep{BenZeev2023}.

\subsection{Setup and simulation regime}
\label{sec:setup}

\begin{figure}[htbp]
\centering
\includegraphics[width=\linewidth]{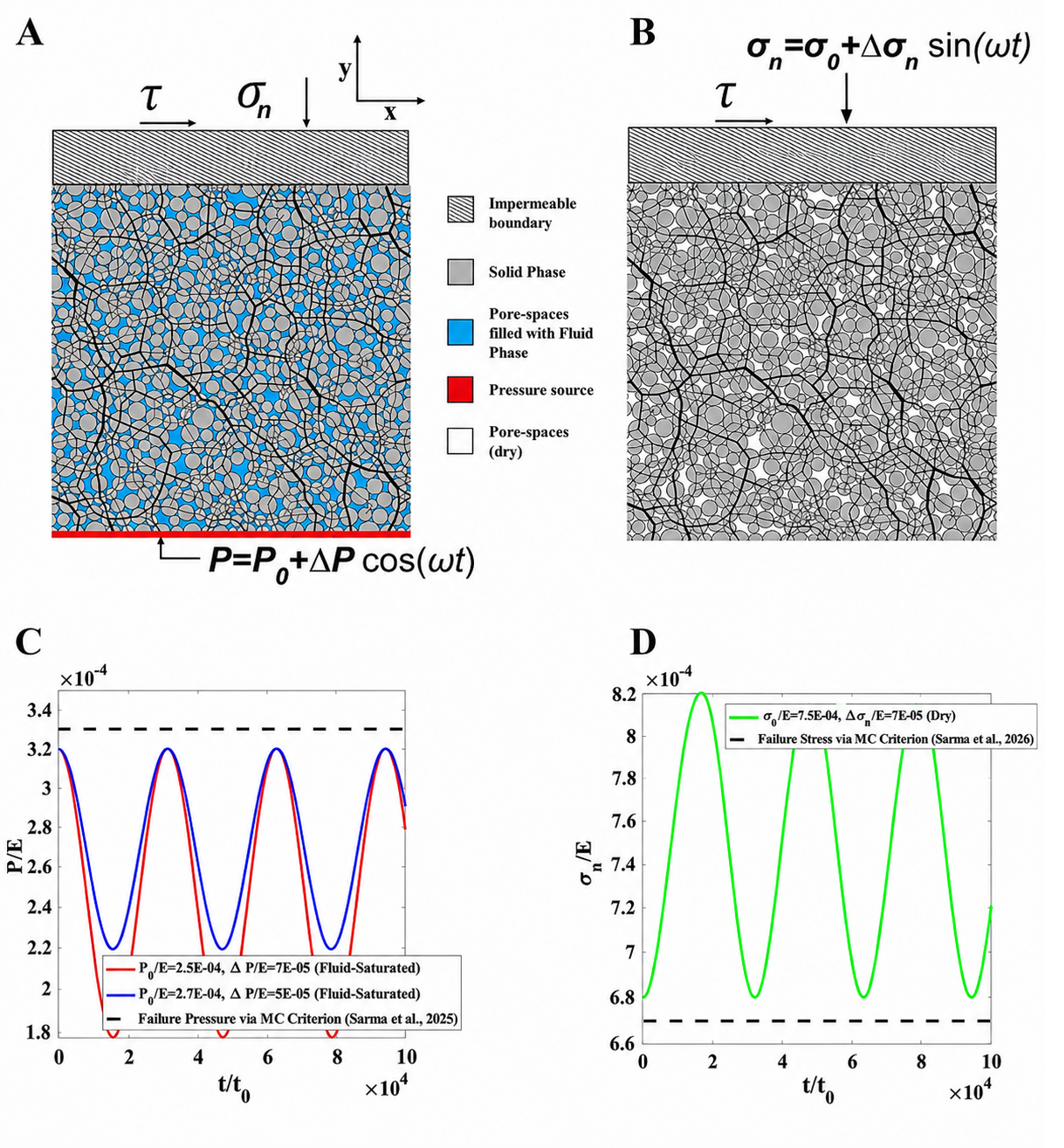}
\caption{Simulation setup and imposed forcing. (A) Dry control (Protocol 1): a prestressed granular layer under constant shear $\tau$ and cyclic normal stress $\sigma_n = \sigma_0 + \Delta\sigma_n\sin(\omega t)$. (B) Fluid-saturated (Protocol 2): an identical layer with impermeable top and cyclic pore pressure $P = P_0 + \Delta P\cos(\omega t)$ at the base; $\tau$ and $\sigma_n$ are fixed and $P$ modulates $\sigma'_n = \sigma_n - P$. (C) Dry forcing $\sigma_n/E$: failure is approached by lowering $\sigma_n$, so the trough stays just above threshold. (D) Wet forcing $P/E$ for the two amplitude pairs: failure is approached by raising $P$, so the peak stays just below threshold. Dashed lines mark the failure threshold \citep{Sarma2025}, never crossed.}
\label{fig:setup}
\end{figure}

The setup represents an idealized two-dimensional fault gouge geometry (Figure~\ref{fig:setup}). The gouge is a granular layer confined in $y$ by two rigid, parallel, rough walls (planar arrays of glued grains). The bottom wall is static; the top wall is free to move horizontally and vertically. We impose a shear stress $\tau$ and a normal stress $\sigma_n$ on the top wall, with periodic boundaries in the $x$ (shear) direction. The aspect ratio is $l/h = 1$.

DEM parameters (Table~\ref{tab:params}) follow \citet{Sarma2025}. The grain Young's modulus $E = 10^{10}$~Pa is slightly lower than typical gouge values but of the same order. The grain size distribution is Gaussian with mean $d = 0.01$~m and standard deviation 0.005~m (polydispersity $\pm 20\%$). The grain size is intentionally larger than natural gouge \citep{Billi2005} to expedite simulations: increasing $d$ and decreasing $E$ raises the collision timescale $t_0 = d\sqrt{\pi\rho_s / 6E} = 3.63 \times 10^{-6}$~s, allowing a larger time step. All variables are non-dimensionalized using $d$, $E$, and $\rho_s$.

\begin{table}[t]
\centering
\caption{Simulation parameters.}
\label{tab:params}
\begin{tabular}{lll}
\toprule
Parameter & Symbol & Value \\
\midrule
Grain density & $\rho_s$ & 2640 kg\,m$^{-3}$ \\
Grain Young's modulus & $E$ & $10^{10}$ Pa \\
Grain mean diameter & $d$ & 0.01 m \\
Grain friction coefficient & $\mu_g$ & 0.5 \\
Mean normal stress & $\sigma_0$ & 10 MPa \\
Shear stress & $\tau$ & $0.2\,\sigma_0$ \\
Fluid density & $\rho_f$ & 1000 kg\,m$^{-3}$ \\
Fluid compressibility & $\beta$ & $10^{-9}$ Pa$^{-1}$ \\
Fluid dynamic viscosity & $\eta$ & $10^{-3}$ Pa\,s \\
Average permeability & $k$ & $10^{-11}$ m$^{2}$ \\
Characteristic time & $t_0 = d\sqrt{\pi\rho_s/6E}$ & $3.63 \times 10^{-6}$ s \\
Characteristic velocity & $V^* = d/t_0$ & $2.75 \times 10^{3}$ m\,s$^{-1}$ \\
Layer thickness & $h$ & $\approx 48\,d$ \\
Initial porosity & $\phi_0$ & $\approx 0.185$ \\
Velocity-strengthening coefficient & $\lambda$ & $\approx 232.5\,t_0$ (wet) \\
\bottomrule
\end{tabular}
\end{table}

Layers are initialized as static dense packings under $\sigma_0 = 10$~MPa with initial shear stress $\tau = 0.2\,\sigma_0$ ($\sim$70\% of static shear strength). The packings are prepared in a dense, over-consolidated state (initial porosity $\phi_0 \approx 0.185$, below the critical-state value), so that the layer must dilate in order to shear and ultimately fail \citep{Reynolds1885, Bolton1986}. Two complementary forcing protocols are then applied:

\textbf{Protocol 1 (control case): direct cyclic $\sigma_n$ oscillations on a dry layer.} No fluid is added (the layer is dry), and the top-wall normal stress oscillates as
\begin{equation}
\sigma_n(t) = \sigma_0 + \Delta\sigma_n \sin(\omega t),
\label{eq:cyclic_sn}
\end{equation}
where $\sigma_0$ is the mean normal stress, $\Delta\sigma_n$ the amplitude, and $\omega = 2\pi f$ the angular frequency. This setup mirrors the experimental geometry of previous normal-stress-oscillation studies \citep{Boettcher2004, Pignalberi2024} and earlier DEM work \citep{Griffa2013}. Note that for normal-stress forcing, failure (and the accompanying dilation) is approached by \emph{reducing} $\sigma_n$, which lowers the effective normal stress; accordingly, the \emph{minimum} instantaneous normal stress $\sigma_0 - \Delta\sigma_n$ is kept just \emph{above} the static failure threshold, so that the imposed $\sigma_n$ never drops to the value that would cause quasi-static failure (Figure~\ref{fig:setup}C). This is the dry-layer counterpart of raising the pore pressure toward, but not above, the failure pressure in the wet case.

\textbf{Protocol 2: cyclic pore-pressure oscillations on a fluid-saturated layer.} The pore space is fluid-filled; the top boundary is impermeable; cyclic pore pressure is imposed at the bottom boundary (Figure~\ref{fig:setup}B):
\begin{equation}
P(t) = P_0 + \Delta P \cos(\omega t),
\label{eq:cyclic_pressure}
\end{equation}
with $\sigma_n$ held constant at $\sigma_0$, so $\sigma'_n(t) = \sigma_0 - P(t)$. Two amplitude combinations are used (Figure~\ref{fig:setup}D): (i) $P_0/\sigma_0 = 0.25$ with $\Delta P/\sigma_0 = 0.07$, and (ii) $P_0/\sigma_0 = 0.27$ with $\Delta P/\sigma_0 = 0.05$. Amplitudes are matched between protocols so that $\Delta\sigma_n = \Delta P$ (non-dimensional). As in Protocol 1, the peak pressure $P_0 + \Delta P$ is kept just below the failure pressure determined from quasi-static pressurization \citep{Sarma2025}.

In both protocols, frequency $f$ is varied from 0.5~Hz to $10^4$~Hz. Comparing the protocols tests directly whether pore-pressure oscillations on a wet layer are equivalent to direct $\sigma_n$ oscillations on a dry layer per Terzaghi's principle. Additionally, comparison simulations at five representative frequencies (1, 10, 100, 1000, $10^4$~Hz) are performed with shear stress set to zero ($\tau = 0$), all other parameters identical, to isolate the purely vibrational component of the response.

\section{Results}
\label{sec:results}

\subsection{Sub-critical failure: three failure modes and one non-failure window}
\label{sec:modes}

Despite peak loading remaining below the quasistatic threshold (Figure~\ref{fig:setup}C, D), the system fails at most frequencies tested under both protocols. The quasistatic threshold (dashed line) was determined from slow, constant-rate pressurization \citep{Sarma2025} using identical initial configuration as in this work. This sub-critical failure is consistent with experimental observations on creeping and stably-sliding gouges \citep{Boettcher2004, Pignalberi2024, Chen2024} and DEM evidence that small dynamic perturbations advance failure \citep{Griffa2013, Ferdowsi2015}.

\begin{figure}[htbp]
\centering
\includegraphics[width=\linewidth]{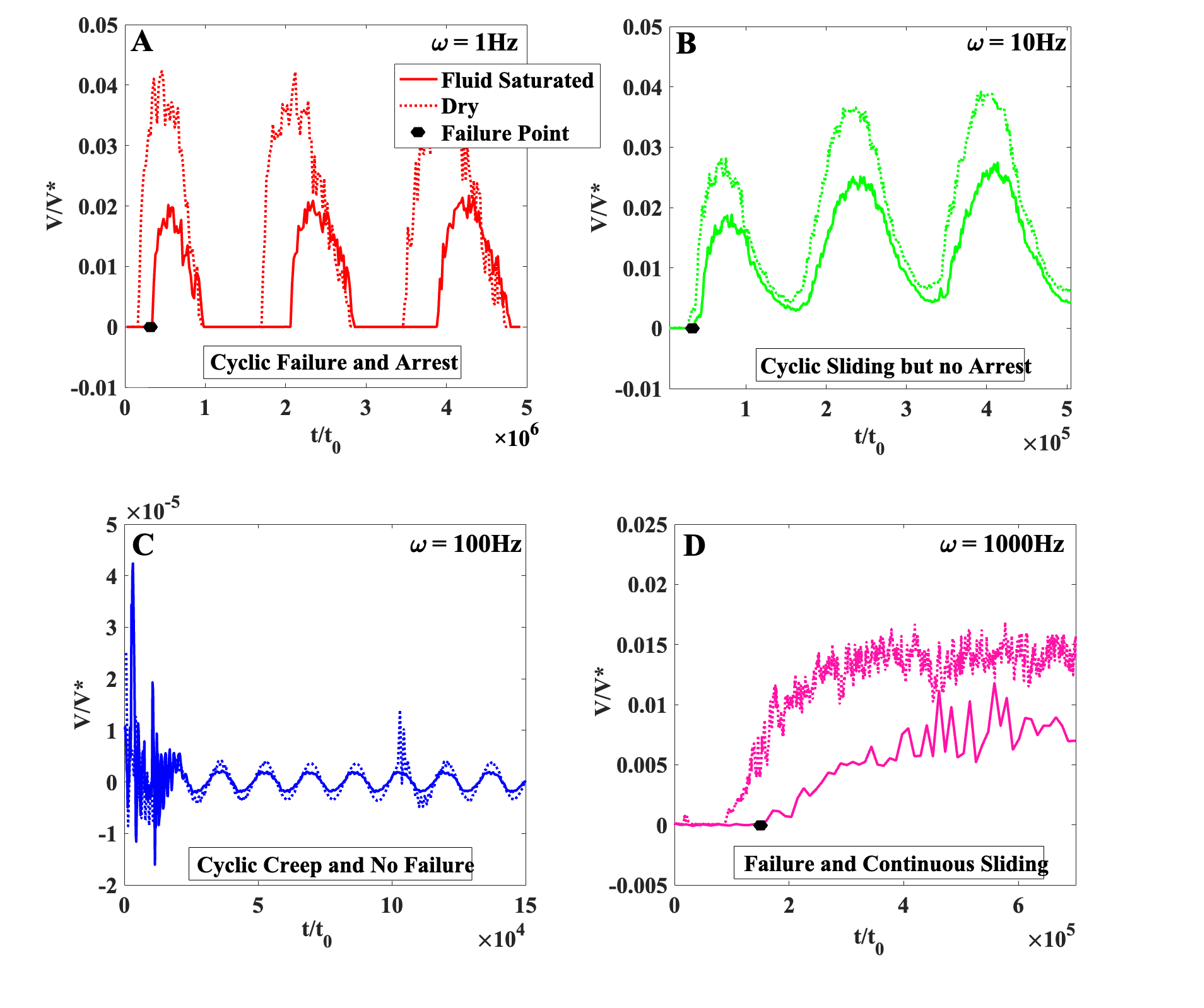}
\caption{Three failure modes and one non-failure window: slip velocity $V/V^*$ versus time $t/t_0$ at four frequencies. Solid: fluid-saturated (Protocol 2); dotted: dry (Protocol 1). Black points mark the failure onset $t_\mathrm{fail}$ (used in Figure~\ref{fig:porosity}); the forcing cycles are in Figure~\ref{fig:setup}C,D. (A) Mode~I: failure and arrest, $\omega = 1$~Hz. (B) Mode~II: sliding without arrest, $\omega = 10$~Hz. (C) Non-failure: cyclic creep, $\omega = 100$~Hz ($\times 10^{-5}$ scale). (D) Mode~III: continuous sliding, $\omega = 1000$~Hz. $\Delta P/\sigma_0 = \Delta\sigma_n/\sigma_0 = 0.07$.}
\label{fig:velocity_modes}
\end{figure}

Throughout, we monitor the horizontal velocity $V$ of the top wall as a function of time, normalized by the characteristic velocity $V^*$ (defined in Table~\ref{tab:params}). The layer is said to \emph{fail} when this velocity exceeds a small threshold and the wall begins to slide macroscopically; we take this failure onset as the reference time $t_\mathrm{fail}$ (marked in Figure~\ref{fig:velocity_modes}, for the wet simulations). Examination of the velocity time series across the full frequency range reveals three distinct \emph{failure} modes (I, II, III) and a separate \emph{non-failure} mode at intermediate frequencies (Figure~\ref{fig:velocity_modes}). Both protocols are shown together: solid lines for the fluid-saturated layer under cyclic pore-pressure forcing (Protocol 2), dotted lines for the dry control under cyclic $\sigma_n$ (Protocol 1). The corresponding imposed forcing cycles are shown in Figure~\ref{fig:setup}C,D.

\begin{figure}[htbp]
\centering
\includegraphics[width=\linewidth]{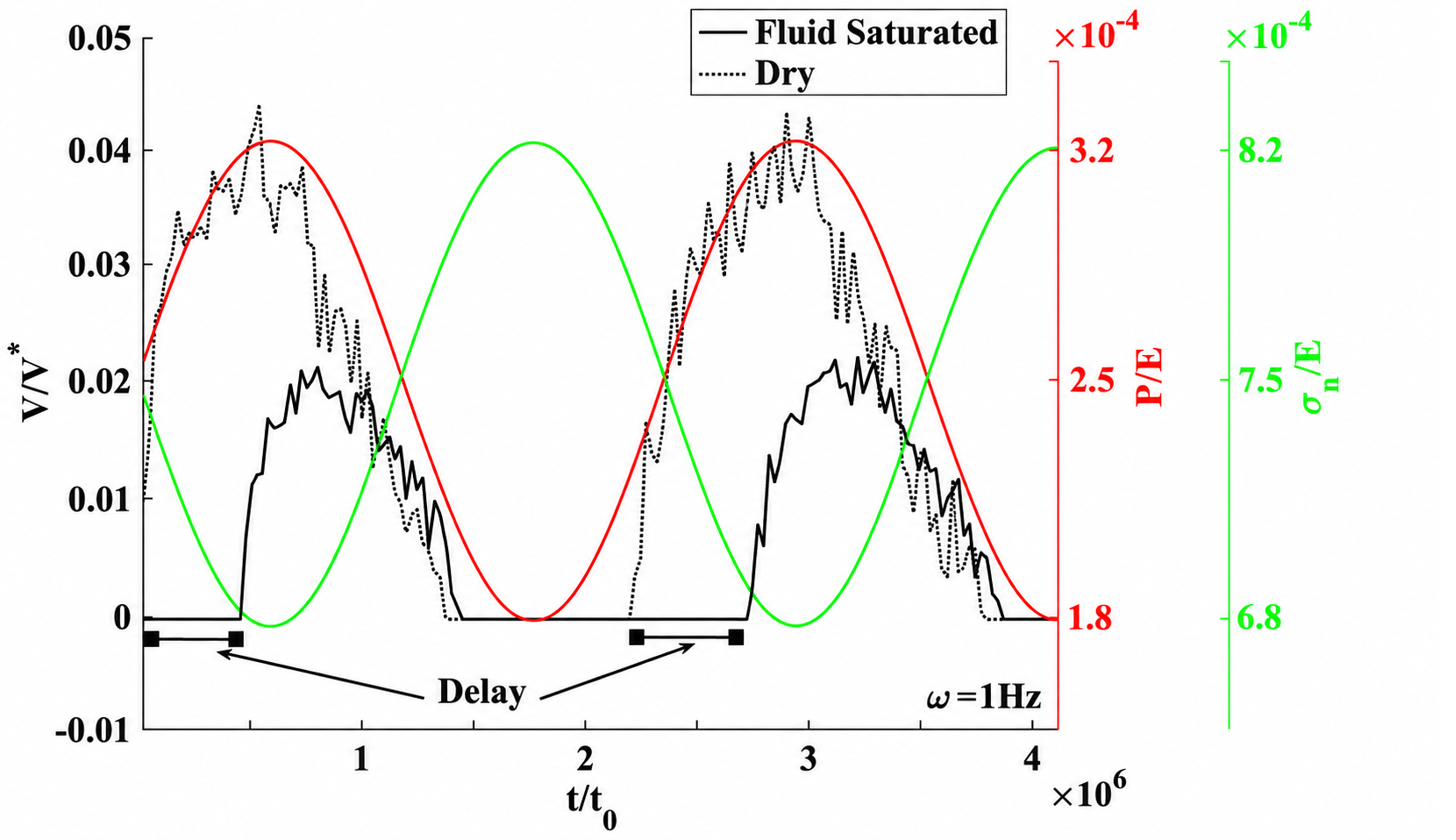}
\caption{Periodicity and delay at $\omega = 1$~Hz. Slip velocity $V/V^*$ for the fluid-saturated (solid) and dry (dotted) layers, overlaid on the imposed forcing: pore pressure $P/E$ (red) and normal stress $\sigma_n/E$ (green). Slip recurs once per forcing cycle; saturated-layer slip is delayed relative to the dry layer by dilatant hardening (marked).}
\label{fig:periodicity_delay}
\end{figure}

\textbf{Mode I: failure and arrest} ($f \sim 0.5$--$1$~Hz). At the lowest frequencies (Figure~\ref{fig:velocity_modes}A), the system fails and arrests in synchrony with the loading cycle, with peak velocities $V/V^* \sim 10^{-2}$. During the high-pressure (low-$\sigma_n$) phase the layer dilates, weakens, and fails; as $\sigma'_n$ is restored it decelerates and arrests, returning to a jammed state, reminiscent of the hysteretic loading--unloading cycles under quasi-static effective-stress variation \citep{Sarma2025}. The wet-case slip is delayed relative to the dry case, which slips at the peak of effective-stress reduction, and its peak slip rate is reduced by fluid drag; both the periodicity of slip and this delay are shown in Figure~\ref{fig:periodicity_delay}.

\textbf{Mode II: Cyclic sliding without arrest} ($f \sim 5$--$10$~Hz). At higher frequencies (Figure~\ref{fig:velocity_modes}B), the system undergoes continuous oscillatory sliding modulated by the cycle but never fully arrests between cycles. This reflects decreasing time available for the layer to decelerate and re-jam between successive peaks. Both protocols produce the same response, with the wet case delayed in phase.

\textbf{Non-failure window: Cyclic creep with no failure} ($f \sim 30$--$200$~Hz). At intermediate frequencies (Figure~\ref{fig:velocity_modes}C), the system exhibits only small-amplitude oscillatory creep with no macroscopic failure. Velocities are orders of magnitude smaller than in the failure modes (note $\times 10^{-5}$ scale), and the system remains essentially jammed. This non-failure window is reproduced in both dry and wet protocols, indicating a fundamental feature of the granular response rather than a process induced by fluid coupling.

\textbf{Mode III: Failure and continuous sliding} ($f \geq 500$~Hz). At sufficiently high frequencies (Figure~\ref{fig:velocity_modes}D), the system fails through a fundamentally different mechanism: after an initial period of apparent stability, it undergoes a transition to sustained sliding that is not strongly modulated by the loading cycle. Failure here is not tied to any single load peak: the instantaneous forcing is identical in every cycle, yet the layer slips only after many cycles, indicating that failure is reached through the gradual, cycle-by-cycle accumulation of dilation (Section~\ref{sec:dilation}) rather than by an individual peak in the loading. 

The dry control and the wet protocol produce the same dynamics at the same frequencies, confirming that the underlying mechanism is the modulation of the effective normal stress. 
The principal quantitative difference is the wet-case delay. In all sliding modes, slip onset in the saturated layers is delayed relative to the dry case. 
This delay is the signature of dilatant hardening characterized in our previous work \citep{Parez2023, Sarma2025}: as the saturated layer dilates in response to reduced $\sigma'_n$, dilation produces a transient pore-pressure drop that temporarily strengthens the layer, delaying macroscopic failure until pore-pressure diffusion restores the pore pressure.

\subsection{Frequency dependence: the non-failure window}
\label{sec:freq_dependence}

\begin{figure}[htbp]
\centering
\includegraphics[width=\linewidth]{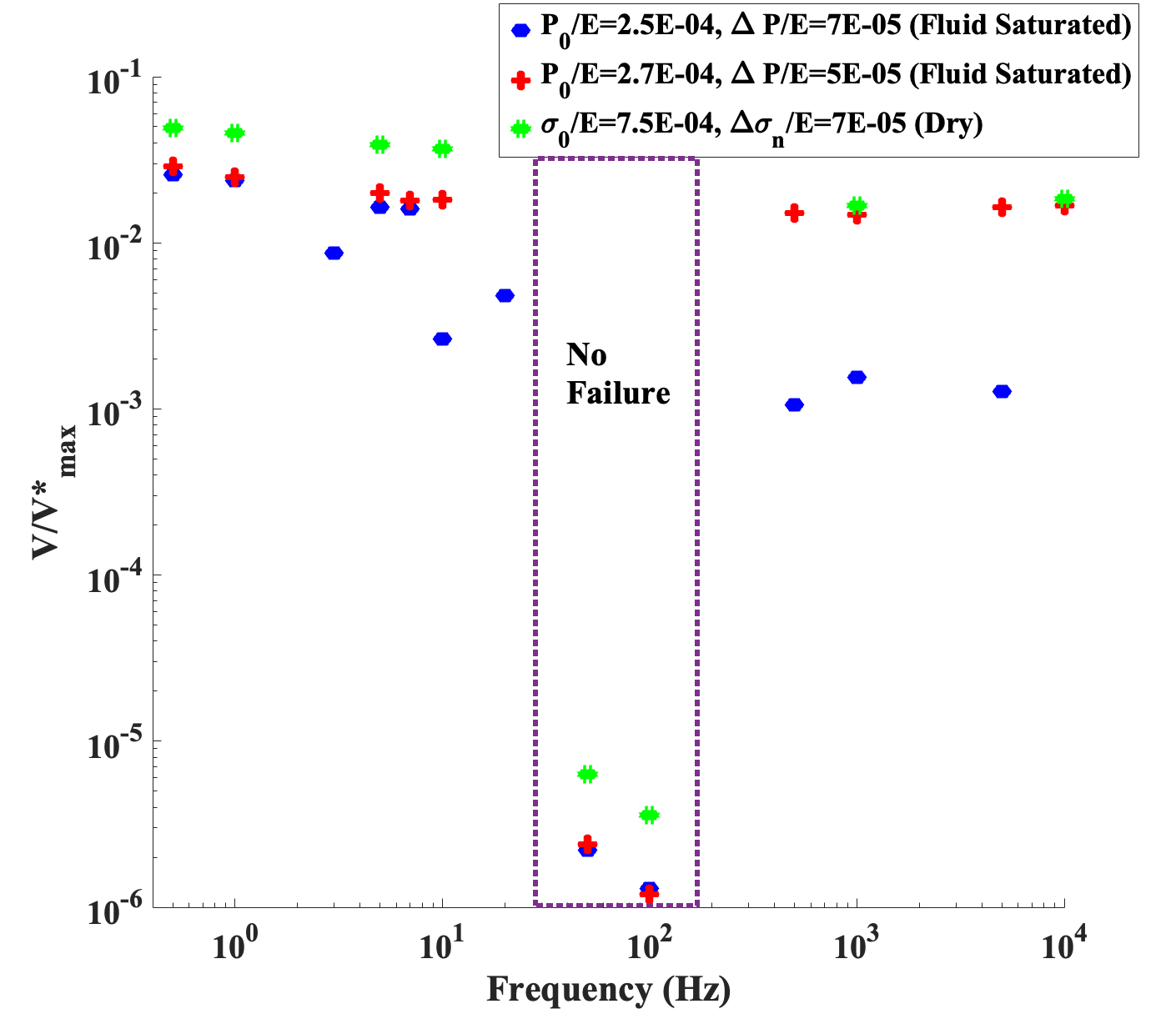}
\caption{Maximum slip velocity $V/V^*_{\max}$ versus oscillation frequency. Fluid-saturated (Protocol 2) at $P_0/\sigma_0 = 0.25$, $\Delta P/\sigma_0 = 0.07$ (blue) and $P_0/\sigma_0 = 0.27$, $\Delta P/\sigma_0 = 0.05$ (red), with the dry control (Protocol 1, $\Delta\sigma_n/\sigma_0 = 0.07$, green). All three share the same non-monotonic dependence, with the minimum in the non-failure window (30--200~Hz, dashed box).}
\label{fig:freq_velocity}
\end{figure}

The maximum normalized slip velocity at each frequency provides a concise summary of frequency-dependent failure (Figure~\ref{fig:freq_velocity}).

At low frequencies (0.5--10~Hz), the system fails consistently with $V/V^*_{\max} \sim 10^{-2}$. As frequency increases, the maximum velocity drops by several orders of magnitude, reaching a minimum in 30--200~Hz (the non-failure window). Beyond 500~Hz, velocity rises again, recovering Mode III. The two pore-pressure perturbations and the dry control all show broadly similar behavior, confirming the non-failure window is a robust feature. The existence of this window cannot be explained by peak-load arguments alone (peak loading is identical across all frequencies); instead, the frequency-dependent behavior arises from the interplay of three physical mechanisms developed below.

\subsection{Dilation as the gateway to failure: ratcheting deterioration of friction}
\label{sec:dilation}

The first mechanism governing failure under cyclic loading is shear-induced \emph{irreversible} dilation. With shear stress $\tau$ held below the static Mohr--Coulomb threshold ($\tau < \mu_s\sigma'_n$), the system can fail only if (i) $\sigma'_n$ drops sufficiently, or (ii) the friction coefficient $\mu$ decreases. Under our sub-critical loading, route (i) is by construction ruled out, so route (ii) must be operative: what physical process causes $\mu$ to deteriorate progressively under cyclic loading?

The answer comes from the porosity dependence of friction, such as that given for example by the Chen--Spiers relation (Equation~\ref{eq:chen_spiers}) \citep{Chen2016}. Because $d\mu(\phi)/d\phi < 0$, any net porosity increase lowers frictional resistance. Densely packed granular media must dilate to shear \citep{Reynolds1885, Reynolds1886, Marone1990, Campbell2006}: each shear-induced rearrangement carries a small porosity increment. Under cyclic loading, the cumulative porosity increase across many cycles drives the layer through progressively weaker $\mu$ values until, at a \emph{failure porosity} $\phi_f$ defined by $\mu(\phi_f)\sigma'_n = \tau$, the available friction can no longer support the imposed shear at the sub-critical effective stress and the layer fails. This failure porosity should be distinguished from the critical-state porosity $\phi_c$ of Equation~\ref{eq:chen_spiers}: since $\mu(\phi_c)$ is the lowest (residual) friction the layer can reach, failure occurs at or below the critical state, $\phi_f \leq \phi_c$, with equality only when the residual friction itself satisfies the failure condition.

\begin{figure}[htbp]
\centering
\includegraphics[width=\linewidth]{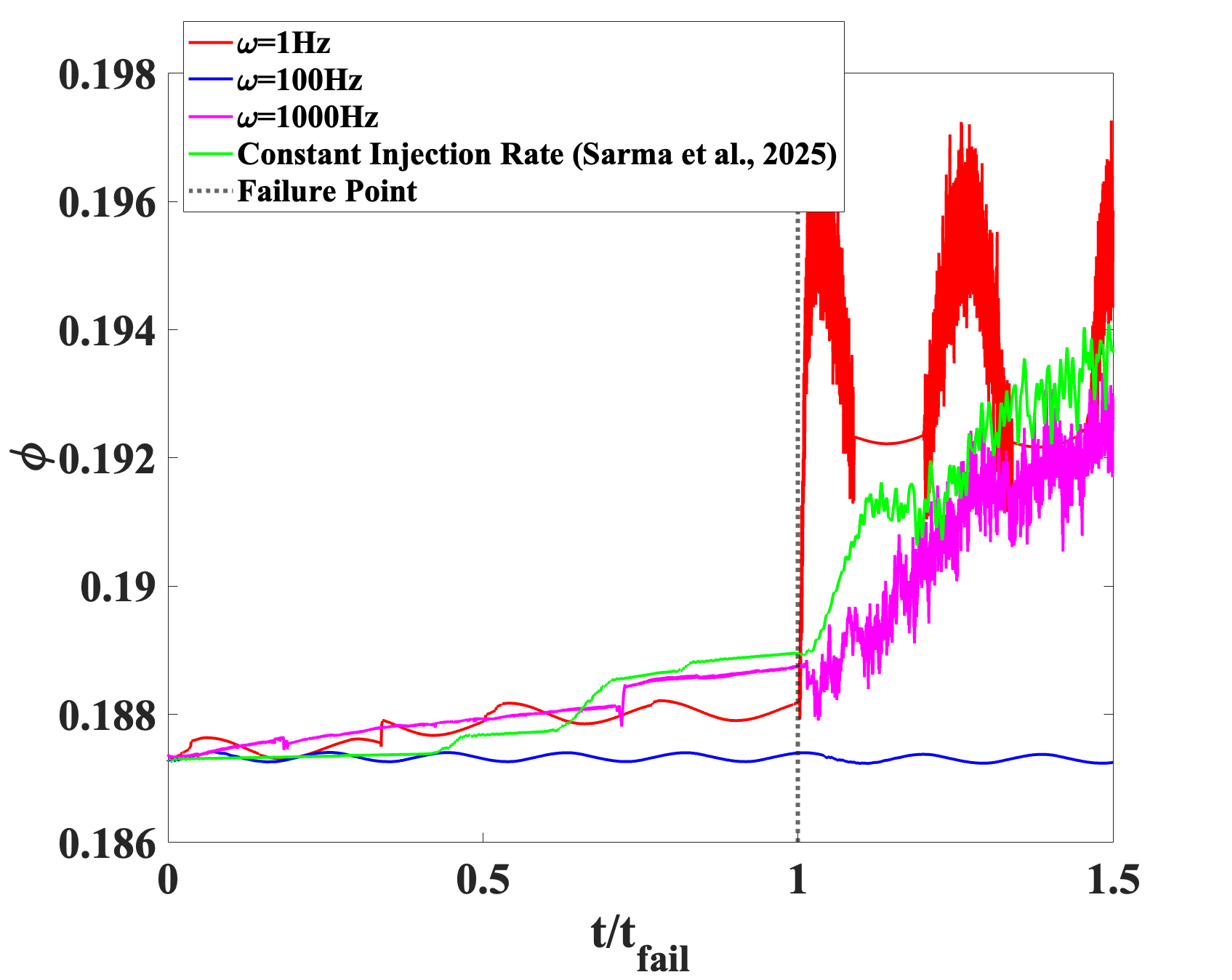}
\caption{Porosity $\phi$ evolution at $\omega = 1$~Hz (red), 100~Hz (blue), 1000~Hz (magenta), and constant injection \citep{Sarma2025} (green). Time is normalized as $t/t_\mathrm{fail}$ so failing runs reach failure at $t/t_\mathrm{fail} = 1$ (onset from Figure~\ref{fig:velocity_modes}). The 1~Hz, 1000~Hz, and constant-injection cases reach the same failure porosity ($\phi_f \approx 0.188$--$0.189$); the 100~Hz case shows negligible net dilation. $P_0/\sigma_0 = 0.25$, $\Delta P/\sigma_0 = 0.07$.}
\label{fig:porosity}
\end{figure}

Figure~\ref{fig:porosity} shows porosity evolution at three representative frequencies (1, 100, 1000~Hz) compared to constant-rate injection \citep{Sarma2025}. The low-frequency (1~Hz) and constant-injection cases both show gradual cumulative porosity increase preceding failure; the 1000~Hz case dilates significantly via a noisier trajectory. All three failing cases reach the same failure porosity range, consistent with a well-defined $\phi_f$ at which $\mu(\phi_f)\sigma'_n = \tau$. The 100~Hz case, in contrast, shows no net dilation: porosity oscillates around its initial value with negligible drift. The lack of cumulative dilation is the proximate reason for stability in the non-failure window.

\begin{figure}[htbp]
\centering
\includegraphics[width=\linewidth]{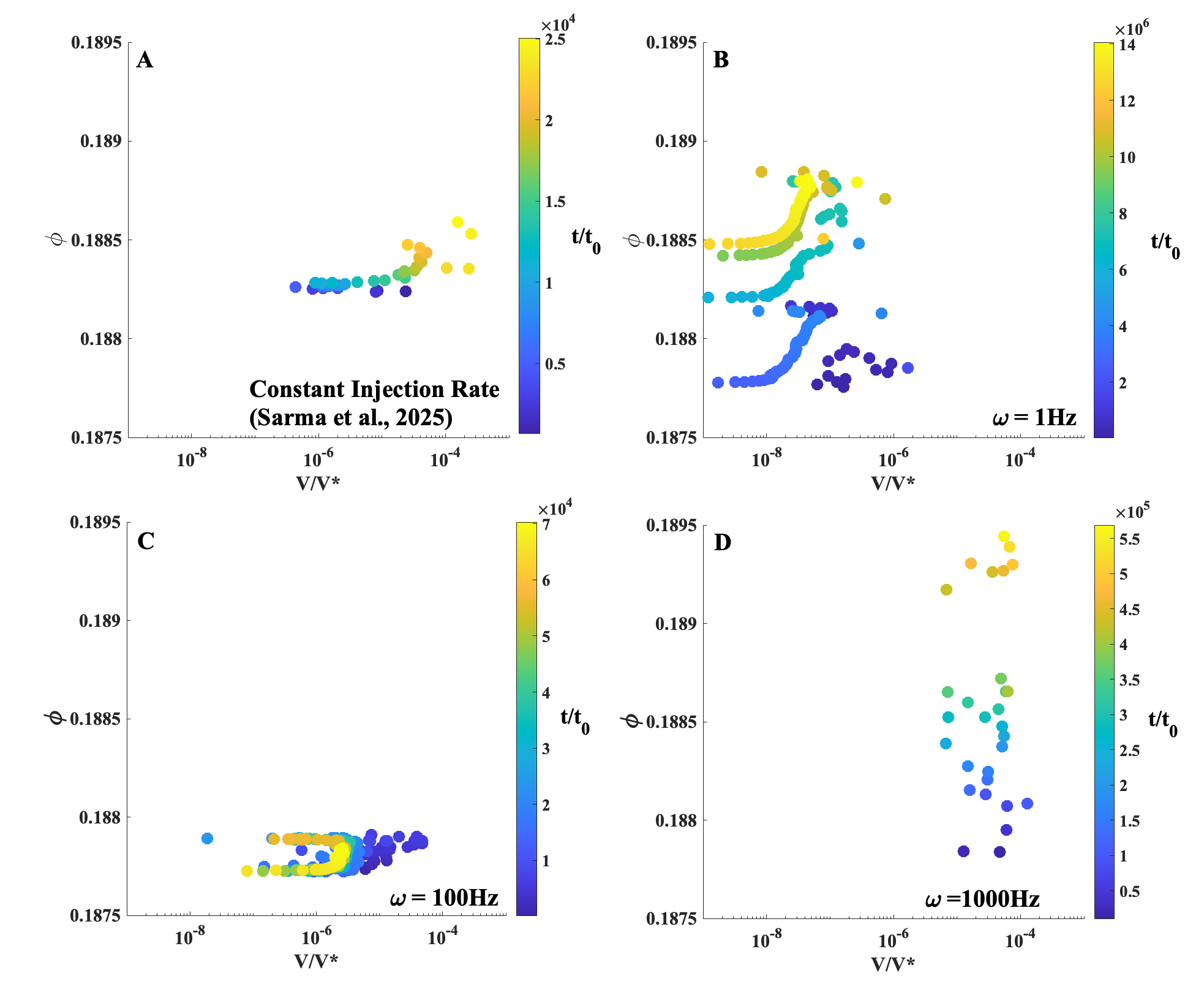}
\caption{Phase-space trajectories of $\phi$ versus $V/V^*$, colour-coded by time, for (A) constant injection \citep{Sarma2025}, (B) $\omega = 1$~Hz, (C) $\omega = 100$~Hz, (D) $\omega = 1000$~Hz. (B) shows the ratchet: each cycle leaves a small irreversible dilation increment (loops drift upward in $\phi$). (C) shows reversible loops with no drift. (D) shows dilation by grain agitation rather than ratcheting. All panels are pre-failure. $P_0/\sigma_0 = 0.25$, $\Delta P/\sigma_0 = 0.07$.}
\label{fig:phi_v}
\end{figure}

Figure~\ref{fig:phi_v} traces the system in $\phi$--$V$ phase space, time-coded by color. The 1~Hz panel (Figure~\ref{fig:phi_v}B) reveals the key mechanism for sub-Coulomb failure: each cycle traces a closed-shape excursion (dilation and acceleration during the high-pressure half-cycle, partial compaction and deceleration during the low-pressure half-cycle), but compaction does \emph{not} fully reverse dilation. Each completed cycle ends at slightly higher porosity, the \emph{ratchet effect}: the cyclic loading produces net positive porosity drift even though pressure returns to the same value at the end of each period. The progressive upward shift of loops in Figure~\ref{fig:phi_v}B (color shifting from dark blue at early times to yellow at late times) makes ratcheting directly visible. Cycle by cycle, $\mu(\phi)$ deteriorates until $\phi_f$ is reached and the layer fails. The constant-injection case (Figure~\ref{fig:phi_v}A) follows a monotonic dilation--acceleration path, the limiting case as the period becomes infinite.

The 100~Hz case (Figure~\ref{fig:phi_v}C) traces small closed loops with no upward drift: each half-cycle's dilation is fully reversed by the next half-cycle's compaction. The ratchet is absent, $\mu(\phi)$ does not deteriorate, and the layer remains stable. The 1000~Hz case (Figure~\ref{fig:phi_v}D) reaches high porosity and high velocity but lacks the clean ratcheting structure of 1~Hz, the first indication that high-frequency dilation is mechanistically distinct from the low-frequency ratchet (further developed in Sections~\ref{sec:forces}--\ref{sec:shear_comparison}).

In summary, sub-Coulomb failure proceeds not by instantaneous overload but by gradual erosion of $\mu(\phi)$: at low frequency through cycle-by-cycle irreversible dilation, and at high frequency through a related but mechanistically distinct dilation process (Sections~\ref{sec:forces}--\ref{sec:shear_comparison}). For a given sliding velocity and grain fabric, any sufficiently low $\sigma'_n$ has a porosity at which $\mu(\phi)\sigma'_n = \tau$; once dilation drives the layer there, failure is inevitable.

\subsection{Competition between momentum diffusion and oscillation timescales}
\label{sec:timescales}

The 100~Hz case does not ratchet and does not fail, while lower and higher frequencies do. The amount of \emph{net} dilation per cycle is controlled by the competition between two timescales: the momentum-diffusion timescale in the granular layer and the imposed oscillation period. In addition, the picture is supplemented at high frequencies by the frequency-dependent magnitude of seepage and inertial forces (Section~\ref{sec:forces}).

Before quantifying this competition, we clarify what \emph{momentum diffusion} means here and why it controls porosity. When the boundary forcing changes, whether $\sigma_n$ is reduced or $P$ is raised, grain motion is not instantaneous everywhere: a velocity, and hence shear-rate, perturbation generated near the forced boundary spreads into the layer over time, governed by a nonlinear diffusion equation for the shear rate \citep{Parez2016, Sarma2025}. This spreading of velocity is the diffusion of momentum. Because the porosity of a sheared granular layer increases with shear rate \citep{Makedonska2011, daCruz2002, MiDi2004}, the diffusing velocity front carries a front of dilation: as momentum reaches a region, grains rearrange and climb over one another, raising the local porosity. Each oscillation of the effective normal stress therefore launches a pulse of momentum, and of dilation, through the layer, and whether this dilation accumulates or is reversed before the next cycle depends on how the momentum-diffusion time compares with the oscillation period. This process is essentially identical in the dry and wet layers, because in the high-permeability regime studied here the pore pressure tracks the boundary forcing almost instantaneously (see below).

\citet{Sarma2025} derived the momentum-diffusion timescale in a velocity-strengthening granular layer:
\begin{equation}
\zeta_v = \frac{4}{\pi^2}\frac{\rho_s (1 - \phi) h^2}{(\sigma_n - P_{\text{inj}})\lambda},
\label{eq:zeta_v}
\end{equation}
where $h$ is layer thickness and $\lambda$ is the strengthening coefficient ($\mu = \mu_{\text{kinetic}} + \lambda\dot{\gamma}$, following $\mu(I)$ rheology). Note that $\zeta_v$ is a property of the grain skeleton alone (set by the grain density, layer thickness, effective normal stress, and strengthening coefficient) and is therefore independent of the pore pressure, and indeed of whether a pore fluid is present at all \citep{Sarma2025}.

In our high-permeability regime, the pore-pressure (hydraulic) diffusion timescale $\zeta_P = \beta\phi\eta h^2 / k$ is substantially \emph{shorter} than $\zeta_v$ \citep{Sarma2025}. The two scale differently ($\zeta_v \propto \rho_s h^2/(\sigma'_n\lambda)$ is governed by grain inertia and frictional strengthening, $\zeta_P$ by fluid storage and permeability), and their ratio $\zeta_P/\zeta_v \sim \beta\phi\eta\,\sigma'_n\lambda/(k\rho_s)$ is small for our high-permeability layer ($k = 10^{-11}$~m$^2$; parameters in Table~\ref{tab:params}). Hydraulic diffusion is therefore the faster process by a wide margin: on the timescale over which momentum redistributes across the layer, the pore pressure has already equilibrated with the imposed boundary pressure $P(t)$, except during rapid pre-failure micro-slip events, whose transient pore-pressure drops (dilatant hardening) produce the wet--dry phase delay in Figure~\ref{fig:velocity_modes}. This makes the dry control mechanically equivalent to the wet protocol at the level of the effective normal stress. 

\paragraph{Loading period vs $\zeta_v$.}
When $T = 1/f \gg \zeta_v$ (low frequencies), each high-pressure half-cycle is long enough for the layer to fully respond: micro-slip-driven dilation accumulates and is only partly reversed by the following compaction, giving a net positive porosity increment per cycle, the ratchet (Figure~\ref{fig:phi_v}B), which over many cycles drives the system to $\phi_f$ and failure (the regime of Modes I and II). When $T \lesssim \zeta_v$, the system cannot fully respond before the load reverses; incipient dilation is essentially fully reversed in the subsequent low-pressure phase, the ratchet vanishes, $\mu(\phi)$ does not deteriorate, and the layer remains stable.

For our parameters (Eq.~\ref{eq:zeta_v}  see Table~\ref{tab:params}), $\zeta_v \sim 10^4\,t_0 \sim 0.04$~s, giving $f_c = 1/\zeta_v \sim 25$~Hz, consistent with the lower edge of the non-failure window in Figure~\ref{fig:freq_velocity}.

The $T \sim \zeta_v$ argument alone, however, cannot explain why the system fails again at high frequencies. The re-emergence of failure at $f \geq 500$~Hz requires a separate mechanism, examined next.

\subsection{Frequency-dependent stress gradients and inertial forces}
\label{sec:forces}

The continuum momentum equation for the granular--fluid mixture, neglecting gravity, is \citep{Goren2010a}
\begin{equation}
\rho \mathbf{v}\cdot\nabla \mathbf{v} + \rho\frac{\partial \mathbf{v}}{\partial t} = -\nabla P - \nabla \cdot \boldsymbol{\sigma}^\prime \,,
\label{eq:mom_full}
\end{equation}
where $\rho$ is the bulk solid density.

Rewriting the convective term as $\nabla(\frac{1}{2}\rho v^2)$ and projecting onto the $y$-direction (normal to the layer)
\begin{equation}
\frac{\partial}{\partial y}\left(\frac{1}{2}\rho v^2\right) + \frac{\partial P}{\partial y} + \frac{\partial \sigma^\prime_{yy}}{\partial y} + \rho\frac{\partial v_y}{\partial t} = 0,
\label{eq:mom_nondim}
\end{equation}
representing, respectively, the kinetic-energy gradient, the seepage force, the effective-stress gradient, and the inertial acceleration. 
%
%

\begin{figure}[htbp]
\centering
\includegraphics[width=\linewidth]{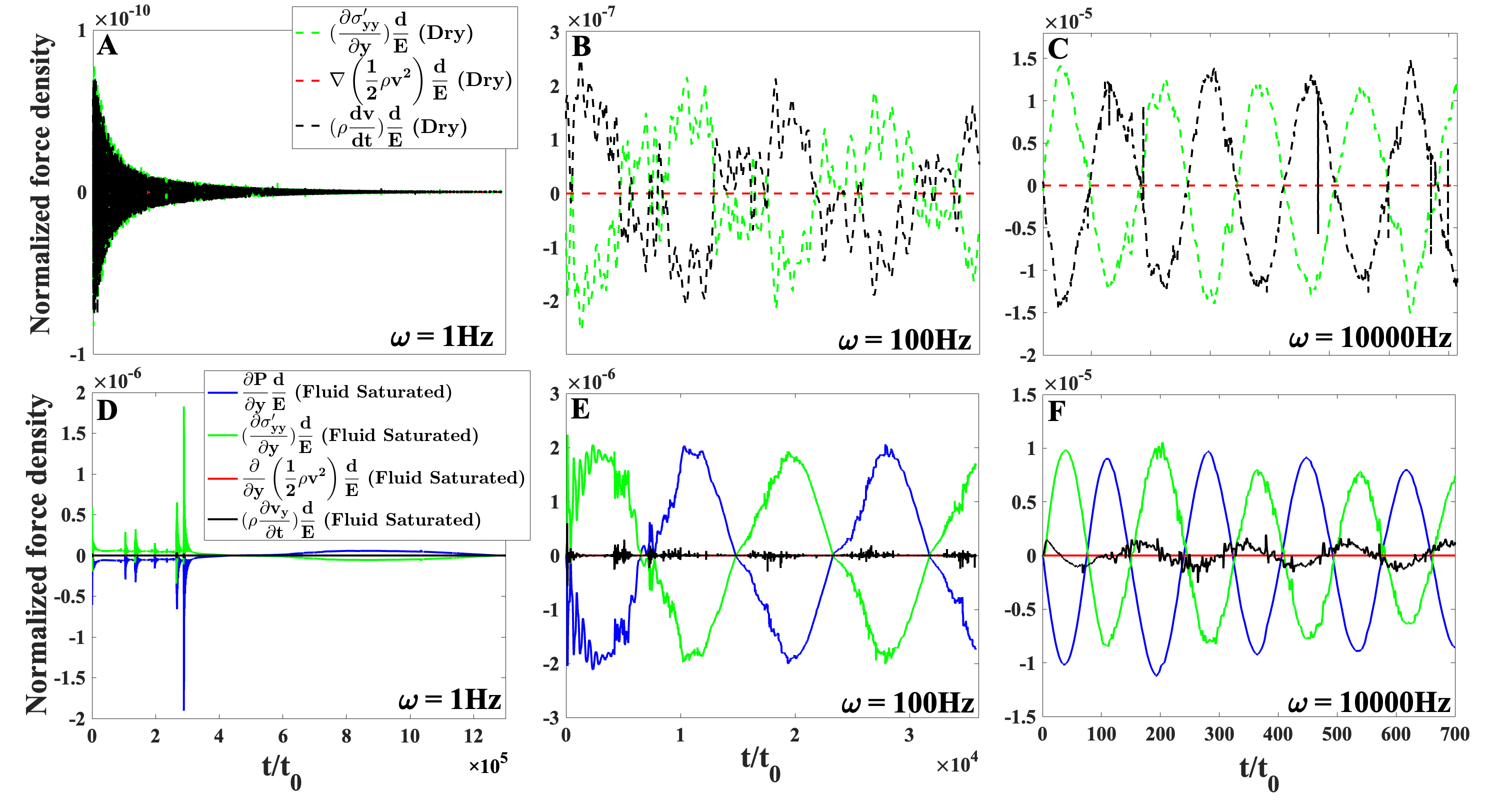}
\caption{Non-dimensional momentum-balance terms (normalized by $d/E$) versus time, Eq.~\ref{eq:mom_nondim}. Top (A--C): dry (Protocol 1) at 1, 100, $10^4$~Hz. Bottom (D--F): wet (Protocol 2) at the same frequencies. Green: solid effective-stress gradient; red: kinetic-energy gradient; black: inertial term; blue (wet only): seepage gradient. Dynamic terms decay at low frequency (A, D); the inertial term grows at high frequency (C, F). Pre-failure phase.}
\label{fig:forces_time}
\end{figure}

Figure~\ref{fig:forces_time} shows time evolution of the three (dry) and four (wet) terms at 1, 100, and $10^4$~Hz. At 1~Hz, the wet pressure-gradient and solid-stress-gradient terms dominate, equal in magnitude and opposite in sign for most of the cycle, so total stress is uniform across the layer, approximate quasi-static equilibrium with the boundary loading, consistent with Section~\ref{sec:timescales}; the dry case behaves the same way (note the small fluctuations). At 100~Hz, the wet seepage and solid-stress gradients oscillate cleanly in anti-phase with comparable magnitudes while the inertial term is small, and in the dry case the stress gradient is balanced by the inertial term. At $10^4$~Hz, the inertial term grows and becomes substantial even in the wet case, confirming the transition into a dynamically dominated regime.

\begin{figure}[htbp]
\centering
\includegraphics[width=\linewidth]{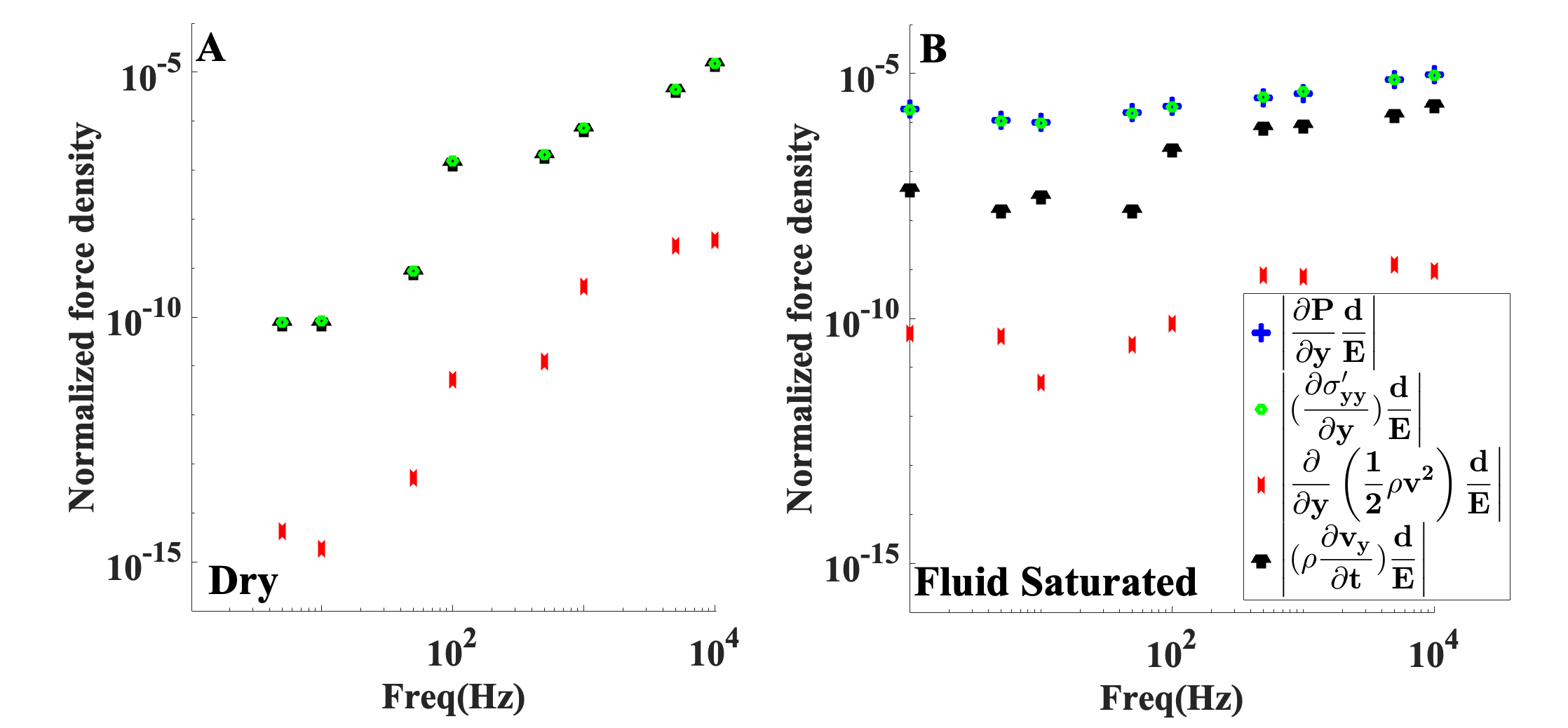}
\caption{Time-averaged magnitudes of the non-dimensional gradient terms (normalized by $d/E$) versus frequency. (A) Dry (Protocol 1); (B) wet (Protocol 2). Colours as in Figure~\ref{fig:forces_time}.}
\label{fig:forces_freq}
\end{figure}

Figure~\ref{fig:forces_freq} summarizes time-averaged force magnitudes versus frequency. In the dry case (A), the solid-stress-gradient and inertial terms are nearly balanced and grow steeply with frequency, while the kinetic-energy-gradient term remains far smaller. In the wet case (B), the pressure gradient and solid-stress gradient increase with frequency and remain nearly equal at all frequencies; the inertial term grows by about two orders of magnitude from 1 to $10^4$~Hz, becoming comparable to the dominant gradients at the highest frequencies.

This reveals the mechanism for high-frequency failure: rapid oscillations generate large dynamic stress gradients (and, in the wet case, pressure gradients) and inertial accelerations whose combined amplified forces agitate the grain assembly, promoting rearrangement and dilation even where quasi-static effective-stress arguments predict stability. Critically, the mechanism does \emph{not} require pore-fluid forcing: the dry control reproduces the same high-frequency failure regime through solid-stress gradients and inertial terms alone. This interpretation is consistent with experiments showing that frictional weakening under vibration is governed by the characteristic vibration velocity \citep{Vidal2019}.

\section{Discussion}
\label{sec:discussion}

\subsection{Two end-member mechanisms; the non-failure window as a gap}
\label{sec:two_mechanisms}

The frequency-dependent failure landscape reflects two distinct end-member mechanisms for achieving dilation, with the non-failure window arising from the gap between them.

\textbf{End-member 1: Shear-driven ratcheting (low frequency).} At $f \leq 10$~Hz, $T \gg \zeta_v$. Each high-pressure half-cycle is long enough for shear-induced micro-slip to drive a dilation increment that the subsequent compaction does not fully reverse: the ratchet (Figure~\ref{fig:phi_v}B). Cycle by cycle, $\mu(\phi)$ deteriorates via Equation~\ref{eq:chen_spiers} until $\phi_f$ is reached. Failure here is fundamentally shear-driven; in the wet case it is paced by the competition between dilation rate and pore-pressure diffusion (dilatant hardening) \citep{Sarma2025}.

\textbf{End-member 2: Vibration-driven failure (high frequency).} At $f \geq 500$~Hz, the oscillation period is much shorter than $\zeta_v$ and the clean, shear-driven per-cycle ratchet is suppressed. Instead, the amplified oscillatory seepage and inertial forces of Section~\ref{sec:forces} agitate the assembly and drive dilation without requiring shear, analogous to vibration-induced fluidization of granular materials \citep{Dijksman2011,Vidal2019} and to dynamic triggering of pre-stressed faults \citep{Ferdowsi2015}. This vibration-driven weakening \emph{is} acoustic fluidization: high-frequency fluctuations transiently relieve inter-grain contact stresses and let the granular mass flow at low macroscopic shear \citep{Melosh1979, Melosh1996, Collins2003}, the mechanism originally invoked for the long runout of rock avalanches \citep{Campbell2006}, with the oscillatory seepage and inertial forces playing the role of the acoustic field. The porosity is \emph{not} drift-free: it still climbs toward failure, but each cycle contributes only a minute increment that accumulates over very many cycles, punctuated by abrupt rearrangement jumps, into the net dilation that triggers failure (Figure~\ref{fig:porosity}).

\textbf{The non-failure window} ($f \sim 30$--$200$~Hz) is a gap between these end-members. The oscillations are too fast for shear-driven ratcheting (insufficient time per half-cycle for dilation to accumulate) but too slow for vibration-driven dilation (oscillatory seepage/stress-gradient forces are not yet large enough to drive grain rearrangement). The system is trapped in a stable state, unable to dilate and consequently fail by either mechanism.

\subsection{Role of shear stress in promoting dilation-induced weakening}
\label{sec:shear_comparison}

\begin{figure}[htbp]
\centering
\includegraphics[width=\linewidth]{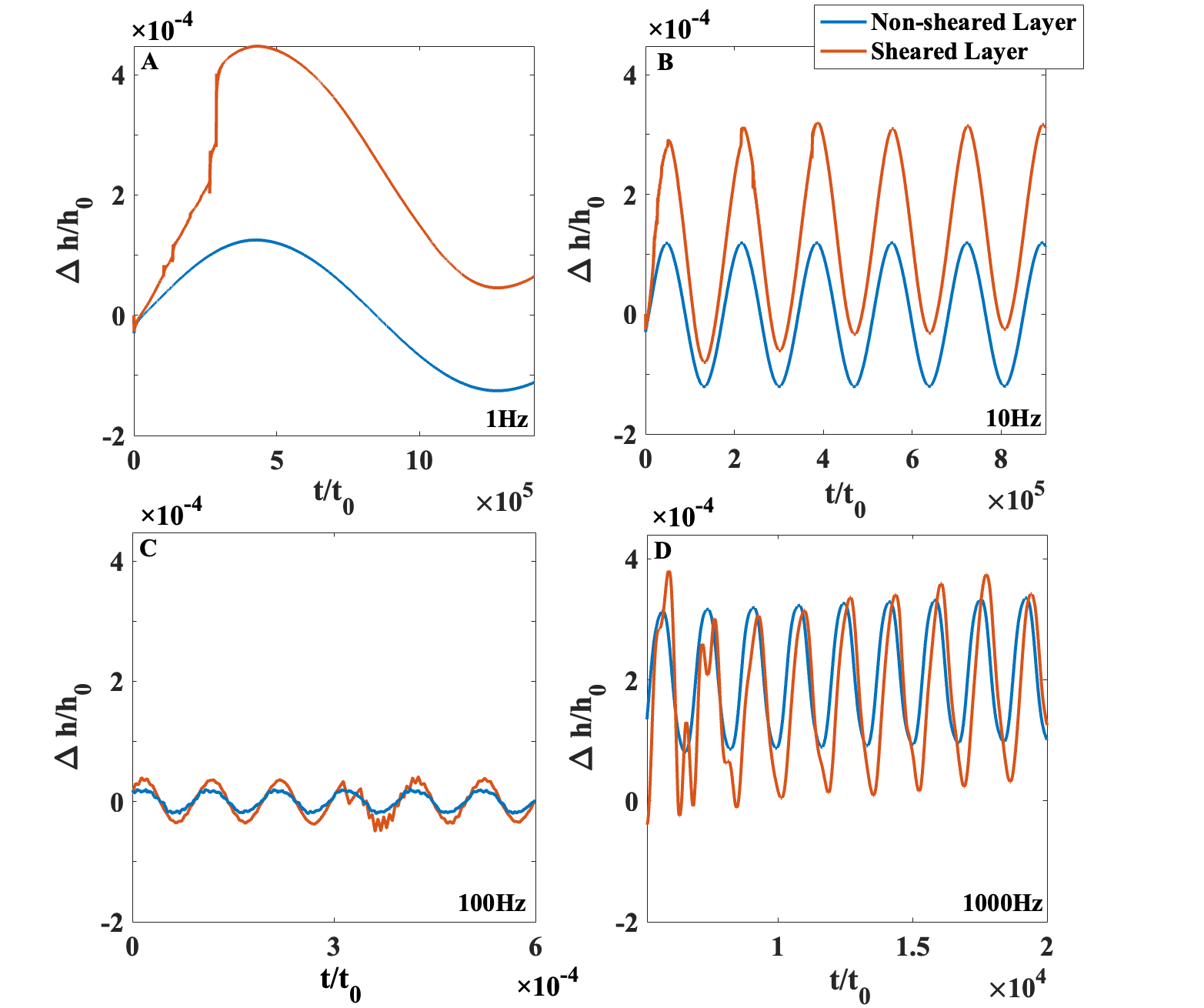}
\caption{Normalized layer-thickness change $\Delta h/h_0$ for sheared ($\tau = 0.2\,\sigma_0$, orange) and non-sheared ($\tau = 0$, blue) layers under cyclic pore pressure at (A) 1~Hz, (B) 10~Hz, (C) 100~Hz, (D) 1000~Hz. (A, B): the sheared layer dilates more on each pressure rise and compacts more on each pressure drop than the non-sheared layer, its minima ratcheting upward, while the non-sheared traces stay symmetric, shear is essential at low frequency. (C): both show negligible net dilation. (D): the two are nearly identical, high-frequency dilation is shear-independent. $P_0/\sigma_0 = 0.25$, $\Delta P/\sigma_0 = 0.07$.}
\label{fig:shear_comparison}
\end{figure}

The two-mechanisms framework makes a testable prediction: removing shear should suppress dilation at low frequencies but have little effect at high frequencies. We test this on the fluid-saturated layer with $\tau = 0$, all other parameters identical to the sheared cases, at four representative frequencies (1, 10, 100, 1000~Hz; Figure~\ref{fig:shear_comparison}).

At low frequencies
(Figure~\ref{fig:shear_comparison}A, B) shear is essential to drive dilation.
At 1~Hz the sheared layer reaches peak dilation $\Delta h/h_0 \approx 4.5\!\times\!10^{-4}$, roughly four times that of the non-sheared layer ($\sim 1\!\times\!10^{-4}$). More importantly, the trace shapes differ: the non-sheared layer expands during high-pressure phases and \emph{compacts back below} its starting thickness during low-pressure phases (symmetric, no net drift), whereas the sheared layer shows successive dilation maxima with cyclic minima shifting progressively upward, irreversible ratcheting toward failure.

At 100~Hz (intermediate frequency; Figure~\ref{fig:shear_comparison}C) both sheared and non-sheared layers show negligible net dilation. The non-failure window is therefore not a property of shear loading but a fundamental feature of the system's response when the oscillation period is comparable to $\zeta_v$.

At 1000~Hz (high frequency; Figure~\ref{fig:shear_comparison}D) the sheared and non-sheared traces show the same high-amplitude pattern with no systematic per-cycle drift: dilation is driven by the oscillatory pressure and inertial forces themselves and proceeds equally well without shear. The few cycles shown are too short to read a net trend; the cumulative high-frequency dilation is instead resolved over the many cycles of Figure~\ref{fig:porosity}, where porosity climbs by grain agitation rather than the cleanly resolved low-frequency ratchet.


\section{Implications for natural and engineered systems}
\label{sec:implications}

Our central finding, that cyclic effective normal stress oscillations induce failure below the Mohr--Coulomb criterion, imposed via either dry $\sigma_n$ variation or saturated pore-pressure modulation, provides a grain-scale explanation for sub-Coulomb failure under oscillatory normal stress \citep{Boettcher2004, Pignalberi2024, Chen2024} and for dynamic-triggering experiments on granular gouge \citep{Johnson2008, Savage2007}. Our DEM-resolved frequency dependence is consistent with rate-and-state friction theory \citep{Linker1992, Dieterich1992} and with the clock-advance phenomenology established by earlier DEM studies \citep{Griffa2013, Ferdowsi2015}. We extend that body of work in three respects: (a) we resolve the frequency dependence over four decades, including conditions under which the system does \emph{not} fail; (b) we identify the underlying grain-scale mechanism (ratcheting irreversible dilation eroding $\mu(\phi)$ at low frequencies, complemented by vibration-driven dilation at high frequencies); and (c) we show by direct dry/wet comparison that, in the high-permeability regime studied here, pore-pressure forcing on a saturated layer is mechanically equivalent to direct $\sigma_n$ forcing on a dry layer, with fluid coupling adding only a delay due to dilatant hardening.

The non-failure window has direct relevance to dynamic earthquake triggering. Seismic waves span a broad frequency range, and dynamic triggering operates not only under high-frequency shaking but also at very low frequencies (for example, by long-period surface waves and slowly-varying transient stresses \citep{Gomberg2005, HillPrejean2007}), so the relevant perturbation spectrum spans both end-member regimes identified here. Our results suggest some frequencies are more effective at triggering slip than others, depending on the interplay between perturbation frequency and the gouge's intrinsic timescales, which may help explain observed variability in dynamic triggering thresholds \citep{vanderElst2010}.

The response is also relevant to cyclic injection protocols for geothermal energy and hydraulic fracturing, where pulsed or oscillatory injection has been proposed to mitigate induced-event magnitude \citep{Zang2019, Kwiatek2019, Hofmann2018}. Our results suggest the oscillation frequency is critical: some frequencies (within the non-failure window) may reduce failure likelihood, while others may paradoxically promote failure at sub-critical pressures, so protocol optimization requires knowing the target formation's intrinsic timescales.

The two mechanisms operate in different settings. Tidal loading ($10^{-5}$--$10^{-4}$~Hz) and slow injection transients favour the dilation-driven ratchet, whereas dynamic triggering by seismic waves (0.1--10~Hz) and high-frequency industrial vibrations may engage both mechanisms, with relative contributions set by permeability, gouge thickness, grain size, and consolidation state. The growth of all force terms with frequency (Figure~\ref{fig:forces_freq}) echoes the importance of inertial effects in rapid granular flows \citep{MiDi2004, Forterre2008}, and the shear-dilation-to-vibration transition may be linked to the quasi-static-to-inertial transition in granular rheology \citep{daCruz2002}. The same physics should apply to cyclic loading of saturated sediments in surface settings such as coastal and submarine slopes.

Because the mode boundaries are set by the granular momentum-diffusion timescale $\zeta_v \propto h^2$ (Equation~\ref{eq:zeta_v}), they are not universal: the non-failure window and the mode transitions should shift to lower frequencies as gouge thickness (and system size) increases \citep{Kuhn2009}. Laboratory and field faults, far thicker than our simulated layer, should thus exhibit the same regime sequence displaced in frequency.

Finally, these conclusions apply to the high-porosity, high-permeability regime simulated here, where hydraulic diffusion is fast ($\zeta_P \ll \zeta_v$) and the dry and wet responses coincide. In low-permeability gouges, pore-pressure diffusion becomes slow relative to the granular timescales, adding a controlling timescale and stronger dilatant hardening \citep{Sarma2025}; whether this shifts or splits the non-failure window remains to be investigated.

\section{Conclusions}
\label{sec:conclusions}

Using coupled fluid-DEM simulations, we investigated the response of granular fault gouge to cyclic effective normal stress oscillations across 0.5--$10^4$~Hz, applying two complementary forcing protocols (direct cyclic $\sigma_n$ on a dry layer as control, and cyclic pore-pressure on a fluid-saturated layer), with peak loading just below the Mohr--Coulomb criterion. Our principal findings are as follows.

(1) Cyclic effective normal stress oscillations induce failure of granular fault gouge even when the instantaneous load never reaches the Mohr--Coulomb threshold, providing a grain-scale mechanistic basis for sub-Coulomb failure under sinusoidal $\sigma_n$ oscillation \citep{Boettcher2004, Pignalberi2024, Chen2024} and acoustically-triggered slip in granular gouge \citep{Johnson2008}.

(2) Three failure modes are identified: cyclic failure-and-arrest (Mode I) at low frequencies; cyclic sliding without arrest (Mode II) at intermediate-low frequencies; failure followed by continuous sliding (Mode III) at high frequencies. A separate non-failure window, small-amplitude cyclic creep, occupies an intermediate-frequency window centered near 100~Hz. The non-failure window is robust across loading amplitudes and across both protocols, although its precise frequency bounds are set by the granular momentum-diffusion timescale and are therefore expected to shift with system size ($\zeta_v \propto h^2$).

(3) The mechanism of sub-Coulomb failure is gradual deterioration of $\mu(\phi)$ via a \emph{ratchet effect}: each cycle leaves a small irreversible dilation increment, so porosity drifts upward cycle by cycle until reaching a failure porosity $\phi_f$ at which the weakened friction can no longer support the imposed shear. This is directly visualized in $\phi$--$V$ phase space at 1~Hz, where successive cycles trace closed loops progressively shifting to higher porosity.

(4) The frequency-dependent behavior arises from two end-member mechanisms with a gap between them. At $f \lesssim 10$~Hz, $T \gg \zeta_v$ (where $T = 1/f$ is the oscillation period and $\zeta_v$ the granular momentum-diffusion timescale, Eq.~\ref{eq:zeta_v}) and shear-induced micro-slip drives a net dilation increment per cycle; removing shear eliminates the ratchet, confirming this end-member is shear-driven. At $f \geq 500$~Hz, rapid oscillations generate large stress gradients and inertial accelerations (and in the wet case, large seepage forces) that agitate the grain skeleton; dilation here proceeds independently of shear and is not resolved cycle-by-cycle: each cycle contributes only a minute porosity increment, and these accumulate over very many cycles, punctuated by abrupt jumps, into the net dilation that triggers failure. The dry control reproduces this regime through solid-stress and inertial terms alone.

(5) Direct comparison of dry and wet protocols shows that the two reproduce the same set of three failure modes plus the non-failure window at the same frequencies. The fundamental mechanism is the modulation of the effective normal stress, exactly as Terzaghi's principle predicts. Fluid coupling adds a single, well-defined modification: a delay due to dilatant hardening, most visible in Mode~I where wet slip is delayed relative to dry. The hierarchy $\zeta_P \ll \zeta_v$ in the high-permeability regime studied here \citep{Sarma2025}, the pore-pressure diffusion timescale $\zeta_P$ being much shorter than the momentum-diffusion timescale $\zeta_v$, ensures pore-pressure equilibrium with the boundary forcing on the relevant timescale, making the dry/wet equivalence clean.

These findings have implications for understanding dynamic earthquake triggering, designing cyclic injection protocols for geothermal energy and CO$_2$ sequestration, and predicting the stability of granular fault gouge under oscillatory loading in natural and engineered systems.


\section*{Acknowledgments}
PS and EA acknowledge the support of the BIRD (Bi-national Israel-US Industrial Development)-GoMed Consortium. RT acknowledges the support of the University of Oslo, the Njord Centre, the CNRS IRP D-FFRACT and the Research Council of Norway through the PoreLab Center of Excellence (project number 262644). SP acknowledges the support of the Johannes Amos Comenius Programme (project No.\ CZ.02.01.01/00/22\_008/0004605).

\section*{Declaration of competing interest}
The authors declare that they have no known competing financial interests or personal relationships that could have appeared to influence the work reported in this paper.

\section*{Data availability}
The simulation data and the data underlying each figure presented in this study are openly available at \href{https://doi.org/10.5281/zenodo.21031756}{https://doi.org/10.5281/zenodo.21031756}. The coupled fluid--DEM simulation software used to generate these data is openly available at \href{https://doi.org/10.5281/zenodo.13765084}{https://doi.org/10.5281/zenodo.13765084}.


\section*{Author contributions}
\textbf{Pritom Sarma:} Conceptualization, Data curation, Formal analysis, Software, Investigation, Validation, Visualization, Writing -- original draft, Writing -- review \& editing.
\textbf{Einat Aharonov:} Conceptualization, Funding acquisition, Methodology, Project administration, Resources, Investigation, Writing -- review \& editing, Supervision.
\textbf{Renaud Toussaint:} Conceptualization, Investigation, Writing -- review \& editing, Supervision.
\textbf{Stanislav Parez:} Conceptualization, Formal analysis, Methodology, Resources, Software, Investigation, Writing -- review \& editing, Supervision.

\bibliographystyle{plainnat}
\bibliography{cyclic_refs}

\end{document}